\documentclass[10pt,twocolumn]{article}
\usepackage[top=2cm, bottom=2cm, left=1.8cm, right=1.8cm]{geometry}

 \usepackage{multirow}
\usepackage{graphicx}
\usepackage{caption}
\usepackage{booktabs}                      
\usepackage{balance}
\usepackage{url}
\usepackage[utf8]{inputenc}
\usepackage[english]{babel}
\usepackage{xspace}
\usepackage{subcaption}
\usepackage{xcolor}
\usepackage[absolute,overlay]{textpos}

\usepackage{color, colortbl}

\definecolor{Gray}{gray}{0.9}

\renewenvironment{thebibliography}[1]{
  \begin{oldthebibliography}{#1}
    \setlength{\itemsep}{0.0em}
    \setlength{\parskip}{0.0em}
}
{
  \end{oldthebibliography}
}

\renewcommand{\footnoterule}{%
  \kern -3pt
  \hrule width 1in
  \kern 2pt
}

\makeatletter
\def\url@leostyle{%
  \@ifundefined{selectfont}{\def\UrlFont{}}%
  {\def\UrlFont{}}%
}
\makeatother
\usepackage[hyphenbreaks]{breakurl}

\definecolor{darkred}{RGB}{153,0,0}
\definecolor{darkblue}{RGB}{0,0,99}
\usepackage[colorlinks=true, linkcolor = darkred,   citecolor = darkred, urlcolor = darkblue]{hyperref}
\newcommand{\one}{({\em i}\/)\xspace}
\newcommand{\two}{({\em ii}\/)\xspace}
\newcommand{\three}{({\em iii}\/)\xspace}

\def\eg{\emph{e.g.}\xspace}

\def\ie{\emph{i.e.}\xspace}
\def\etal{\emph{et al.}\xspace}
\def\vs{\emph{vs.}\xspace}

\newcommand{\pb}[1]{\vspace{0.75ex}\noindent{\bf \em #1}\hspace*{.3em}}
\newcommand{\paragraphbe}[1]{\vspace{0.75ex}\noindent{\bf \em #1}\hspace*{.3em}}

\newif
\ifcomment
\commenttrue

\ifcomment
\newcommand\gareth[1]{\textbf{\textcolor{red}{GT: #1}}	}
\newcommand\ar[1]{\textbf{\textcolor{blue}{AR: #1}}	}

\newcommand\ishaku[1]{\textbf{\textcolor{purple}{IA: #1}}	}
\newcommand\ignacio[1]{\textbf{\textcolor{green}{IC: #1}}	}
\newcommand\edc[1]{\textbf{\textcolor{brown}{EDC: #1}}	}

\else
\newcommand\gareth[1]{}
\newcommand\ar[1]{}
\newcommand\ishaku[1]{}
\newcommand\ignacio[1]{}
\newcommand\edc[1]{}
\fi

\colorlet{sharks}{blue}

\begin{document}

\sloppy

\graphicspath{{pleroma_plots/imc_plots}}

\title{Will Admins Cope?  Decentralized Moderation in the Fediverse}
%
\date{}

\author{\normalsize Ishaku Hassan Anaobi$^1$, Aravindh Raman$^2$, Ignacio Castro$^1$, Haris Bin Zia$^1$, Dami Ibosiola$^1$, and Gareth Tyson$^{1,3}$\\[0.5ex]
\small
$^1$Queen Mary University of London, $^2$Telefonica Research, $^3$\hspace{-0.05cm}Hong Kong University of Science and Technology (GZ)\\}

\maketitle

\begin{abstract}

As an alternative to Twitter and other centralized social networks, the Fediverse is growing in popularity.
The recent, and polemical, takeover of Twitter by Elon Musk has exacerbated this trend.
The Fediverse includes a growing number of decentralized social networks, such as Pleroma or Mastodon, that share the same subscription protocol (ActivityPub). 
Each of these decentralized social networks is composed of independent instances that are run by different administrators.
Users, however, can interact with other users across the Fediverse regardless of the instance they are signed up to.
The growing user base of the Fediverse creates key challenges for the administrators, who may experience a growing burden.
In this paper, we explore how large that overhead is, and whether there are solutions to alleviate the burden.
We study the overhead of moderation on the administrators. We observe a diversity of administrator strategies, with evidence that administrators on larger instances struggle to find sufficient resources. We then propose a tool, WatchGen, to semi-automate the process.

\end{abstract}

\section{Introduction}
\label{sec:intro}

The Fediverse encompasses a group of increasingly popular platforms and technologies that seek to provide greater transparency and openness on the web.~\cite{guidi2018managing, lens, inappropriate, dtube}.
Well known Fediverse platforms include microblogging services (\eg Pleroma~\cite{pleroma}, Mastodon~\cite{mastodonurl}) and 
video sharing platforms (\eg PeerTube~\cite{peertubeurl}).
The acquisition of Twitter by Elon Musk~\cite{musk} has exacerbated this popularity with a large migration of Twitter users to the Fediverse~\cite{mastomigration}.

In Fediverse social networks, individuals or organisations can install, own, and manage their own independent servers, also known as \textbf{instances}~\cite{verge17, zignani2018follow}.
For these instances to interact, they rely on  \textbf{federation}~\cite{schwittmann2013sonet}, whereby instances interconnect in a peer-to-peer fashion to exchange posts. Note that this allows for users to exchange content across platforms. This results in a physically decentralized model that is logically interconnected where users can interact globally. 
Unfortunately, this creates challenges for instance \textbf{administrators}, as activities  on one instance  impact others via federation. 
For example, recent work has shown that hateful material generated on one instance can rapidly spread to others~\cite{zia2022toxicity}. 

To overcome this, most Fediverse social network implementations have  in-built \textbf{federation policies}. 
These policies enable administrators to create rules to ban or modify content from instances that matches certain rules, \eg banning content from a particular instance or associating it with warning tags. 
Although a powerful tool, this imposes an additional overhead on administrators~\cite{collusion, communities, quick}. 
Thus, we argue it is vital to better understand this process, and propose ways to improve it.

This paper examines administrator activities in the Fediverse. We focus on Pleroma, a federated microblogging platform with similar functionality to Twitter. We collect a large-scale dataset covering 10 months: this includes 1,740 instances, 133.8k users, 29.9m posts, associated metadata, and importantly, the policies setup by the administrators. 
We find that instances are often ``understaffed'', with the majority of instances only having a single administrator, and recruiting no other moderators to assist, despite many having over 100K posts.
This leads us to conjecture that some administrators may be overwhelmed. 
Indeed, we find that instance administrators often take many months before applying policies against other instances, even in cases where they exhibit clearly controversial traits (\eg posting a large number of hate words).

We therefore turn our attention to the policy configurations employed.
We observe a growing number of instances enacting a wide range of policy types. Common are `maintenance' policies, such as those which automatically delete older posts (\texttt{ObjectAgePolicy}), as well as those aimed at preventing the spread of certain content (\eg \texttt{HashtagPolicy}, which flags up posts with certain hashtags).
We further observe a range of bespoke policies created by administrators, via the \texttt{SimplePolicy}, which can be configured to trigger a range of actions based on certain rules (\eg blocking all connections from certain instances).
The laborious nature of this moderation work leads us to explore automated techniques to assist administrators. We build a set of models to predict administrator actions. We embed them in WatchGen,  a tool that can propose a set of instances for administrators to focus their moderation efforts on.
To the best of our knowledge, this is the first study of Fediverse administrators.
We make the following observations:

\begin{enumerate}
\item We find a diverse range of 49 policies used by administrators, capable of performing various management and moderation tasks. Despite this, we see that 66.9\% of instances are still running, exclusively, on the default policies alone (Section~\ref{section:pol}).

\item The number of administrators does not grow proportionately with the number of posts (Section~\ref{section:admin}). This seems to impact moderation. For example, it takes an average of 82.3 days for an administrator to impose a policy against an instance after it first encounters it, even for well-known and highly controversial ones (\eg~\texttt{gab.com}~\cite{arnold2021moving}).

\item Intuitive features, such as the number of mentions and  frequent use of hate words, are good indicators that an instance will later have a policy applied against it (Section~\ref{sec:watchlist}). This suggests that there are key traits that garner more attention by moderators.

\item We show that it is possible to predict (F1=0.77) which instances will have policies applied against them (Section~\ref{sec:watchlist}) and design \emph{WatchGen}, a tool that flags particular instances for administrators to pay special  attention to.

\end{enumerate}

\section{Pleroma: Overview}
\label{sec:overview}
Pleroma is a lightweight decentralized microblogging server implementation with user-facing functionality similar to that of Twitter.
In contrast to a centralized social network, Pleroma is a federation  of multiple independently operated servers (aka instances). Users can register accounts on these instances and share posts with other users on the same instance, or on different instances.
Through these instances, users are able to register accounts and share posts (called statuses) to other users on the same instance, other Pleroma instances, or instances from other Fediverse platforms, most notably Mastodon.

\pb{Federation.}
We refer to users registered on the same instance as \textbf{local}, and users on different instances as \textbf{remote}. A user on one instance can follow another user on a separate instance. 
Note that a user registered on their local instance does not need to register with the remote instance to follow the remote user.
When the user wants to follow a user on a remote instance, the local instance subscribes to the remote user on behalf of the local user using an underlying subscription protocol (ActivityPub~\cite{activitypub}). This process of peering between instances in the Fediverse is referred to as \textbf{federation}.

The federated network includes instances from Pleroma and other platforms (\eg  Mastodon) that support the same subscription protocol (ActivityPub). Accordingly, Pleroma instances can federate and target their policies at non-Pleroma instances.
The resulting network of federated instances is referred to as the \emph{Fediverse} (with over 23k servers~\cite{fedinfo}).

\pb{Policies.}
Policies affect how instances federate with each other through 
different rule-action pairs. These allow certain actions to be executed when a post, user, or instance matches pre-specified criteria. For example, the \texttt{SimplePolicy} can perform a range of actions when a remote instance matches certain criteria such as \texttt{reject}ing connections. Note, there are numerous in-built policies, but tech-savvy administrators can also write their own bespoke policies.

\pb{Administrators.}
Instances are hosted and managed by specialized users called administrators. 
By default, the creator of an instance will take on the role of the administrator, however, it is also possible to delegate such responsibilities to multiple others.
Instance administrators are responsible for carrying out the day-to-day administrative tasks on the instances.
These include managing the front-end, users, uploads, database, emoji packs and carrying out administrative email tasks. The instance administrator is also responsible for accepting new user registrations and removing users where necessary. The administrator updates and backs-up the instance, set the terms of service and retains the ability to shutdown the instance. One essential responsibility of the instance administrator is the moderation of content (although they can also assign the role to other users called \emph{moderators}). This can make instance administration a cumbersome task, and administrators a very important part of the Fediverse.

\section{Data Collection}
\label{section:data}

\paragraphbe{Instance \& Administrator Dataset.} 
Our measurement campaign covers 16th Dec 2020 -- 19th Oct 2021.
We first compile a list of Pleroma instances by crawling the directory of instances from ~\texttt{\url{distsn.org}} and~\texttt{\url{the-federation.info}}. 
We then capture the list of instances 
that each Pleroma instance has ever federated with
using each instance's Peers API.\footnote{\url{<instance.uri>/api/v1/instance/peers}}
Note, this includes both Pleroma and non-Pleroma instances.
In total, we identify 9,981 instances, out of which 2,407 are Pleroma and the remainder are non-Pleroma (\eg Mastodon).  

We then collect  metadata for each Pleroma instance every 4 hours via their public API.\footnote{\url{<instance.uri>/api/v1/instance/}}
We record the list of administrators and any delegated moderators.
We also obtain the number of users on the instance, the number of posts, the enabled policies, the applied policies as well as the instances targeted by these policies, and other meta information. 

From the 2,407 Pleroma instances, we are able to gather data from a total of 1,740 instances (72.28\%). 
For the remaining 667 instances: 65.1\% have non existent domains, 17.9\% are not found (404 status code), 6.4\% instances has private timelines (403), 4.5\% result in Bad Gateway (502), 1.3\% in Service Unavailable (503), and under 1\% return Gone (410).

\pb{User Timelines.}  Users in Pleroma have three timelines: 
\one~a \textit{home} timeline, with posts published by the accounts that the user follows (local and remote);
\two~a \textit{public} timeline, with all the posts generated within the local instance; 
and \three~the \textit{whole known network}, with \emph{all} posts that have been retrieved from remote instances that the local users follow. 
Note, the \textit{whole known network} is not limited to remote posts that a particular user follows: it is the union of remote posts retrieved by all users on the instance. 
We use the public Timeline API\footnote{\url{<instance.uri>/api/v1/timelines/public?local=true}} to gather posts data from 819 instances (the remaining 912 instances  have either no posts or unreachable public timelines).

\pb{Ethics.}
Our dataset covers Pleroma instances and their administrators.
We exclusively focus on the policies that these administrators set, and do not investigate other aspects of administrator behavior (\eg the posts they share). 
All data is available via public APIs. 
We emphasize that administrators, themselves, are the ones who control access to these APIs. 
Hence, the administrators covered in this paper consent for others to use this data.
Further, the policies studied do not work on a per-user granularity and, thus, we cannot infer anything about individual users. 
All data is anonymized before usage, and it is stored within a secure
silo.

\section{Exploring Policy Configurations} 
\label{section:pol}

\begin{figure}[t]
  \includegraphics[width=\linewidth]{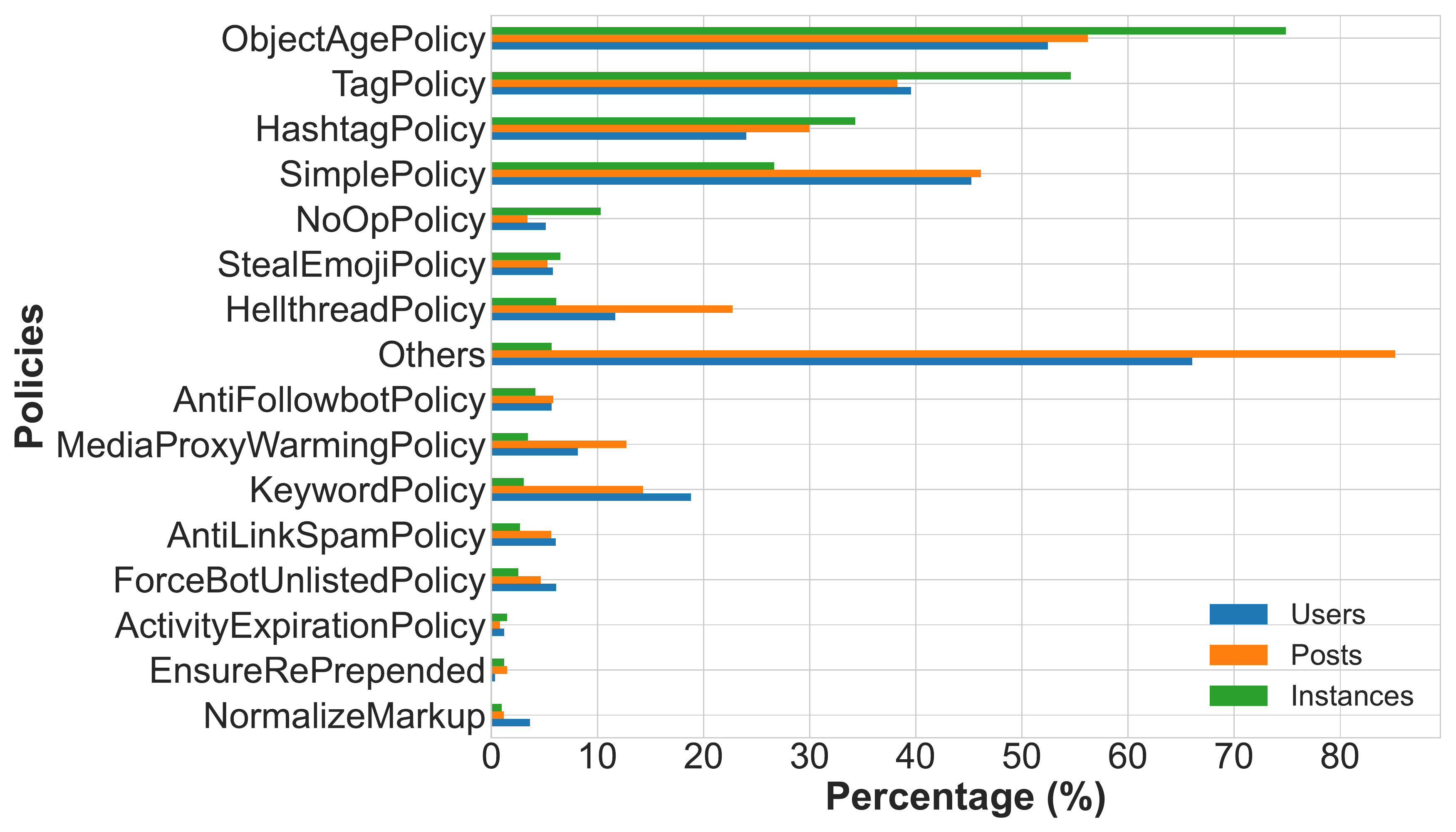}
  \caption{The top 15 policies  and  percentage of instances that use each policy (sorted by the percentage of instances).}
  \label{fig:policies}
\end{figure}

\pb{Policy Footprint.}
We first quantify the presence of policies across instances.
In total, we observe 49 unique policy types.
From our 1.74k Pleroma instances, we retrieve policy information from 93.2\% of instances (the remainder do not expose their policies).
These cover 94.2\% of the total users and 94.5\% of all posts.
Figure~\ref{fig:policies} shows the distribution of the top 15 policy types enabled by the administrators across instances and the percentage of users signed up within those instances as well as the posts on the instances. 
We see a wide range of policies with diverse functionalities and varying coverage based on which metric is considered. For instance, whereas the \texttt{ObjectAgePolicy} (which performs an action on a post once it reaches a certain age) is installed on 74.8\% of instances, this only covers 52.4\% of the users. In contrast, the \texttt{KeyWordPolicy} (which performs an action on any posts containing a given keyword) 
covers 18.8\% of users, but just 3\% instances.
Critically, there is a highly uneven distribution of policies, with the the top-5 covering 92.3\% of all instances, 73.6\% of
users and 88.8\% of the posts.

\pb{Default Policies.} 
Default policies come auto-enabled with new installations. Prior to version 2.3.0 in March, 2021, only the \texttt{ObjectAgepolicy and NoOpPolicy} are enabled by default. Since version 2.3.0, the \texttt{TagPolicy} and \texttt{HashtagPolicy} are also enabled with a new installation (or upgrade).
66.9\% of instances only have these default policies running.
Relying solely on default policies may indicate several things. 
For example, administrators maybe unaware of management and moderation functionalities, unable to use them or simply not have sufficient time. 
Alternatively, they may actively choose not to use them.

Note, while the \texttt{TagPolicy} allows tagging user posts as sensitive (default: nsfw), the \texttt{Hashtagpolicy} allows the tagging of hashtags (\eg nsfw sensitive). We find 54.6\% and 34.3\% of instances enabling these policies respectively. The other Pleroma default policy is the \texttt{NoOpPolicy}. This allows any content to be imported. This describes the default state of a new instance. Interestingly, we see administrators paying more attention to this policy: 89.7\% of the instances have actively disabled it.\footnote{Note, this is overridden if a user enabled any other policy.} 
This suggests that administrators are aware and concerned about importing undesirable content.

\pb{Non-Default Policies.}
Non-default policies are those that instance administrators have to actively enable.
Instances with these policies may indicate a more proactive administrator.
We find 45 non-default policies during our data collection period. 

The most powerful policy available is the  \texttt{SimplePolicy}, enabled on 28.8\% of instances.
This policy allows administrators to apply a wide range of actions against specific instances (\eg \texttt{gab.com}). The most impactful and common is the \texttt{reject} action.\footnote{This blocks all connections from a given instance}
56.9\% of instance that enable the \texttt{SimplePolicy} employ the \textit{reject} action.
Interestingly, although we see only 28.8\% of instances with the \texttt{SimplePolicy} enabled, its application affects 85.4\% of users and 90.3\% of the posts on the Pleroma platform. 
We see noteworthy instances being amongst the top targets of this policy (\eg \textit{kiwifarms.cc and anime.website}), which are all commonly understood to share controversial material.
Interestingly, only 18.5\% of instances with the \texttt{SimplePolicy} applied against them are from the Pleroma platform (the most are from Mastodon~\cite{Mastodon}). This means that 81.5\% of the recipients are from federated instances outside of Pleroma.

\begin{figure}[t]
 \centering
  \includegraphics[width=\linewidth]{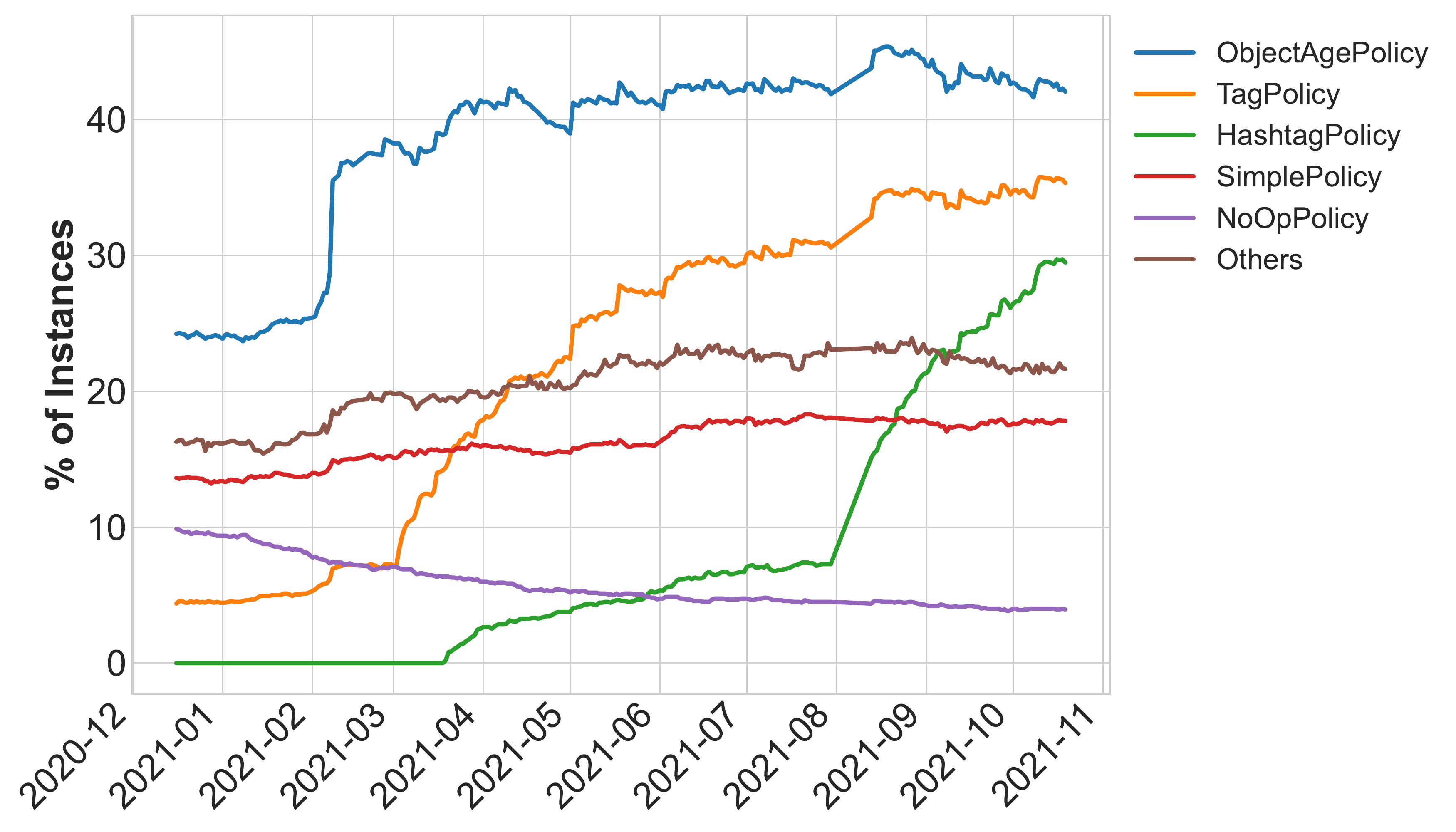}
  \caption{Time series showing the percentage of instances (Y-axis) that use the 5 most popular Pleroma policies. We include the sum of all the remaining policies as ``Others''.}
  \label{fig:policy time}
\end{figure}

\pb{Policy Growth.}
We next look at how the use of policies has changed over time. 
We conjecture that the longer administrators run their instances, the more experienced they become. As such, we expect to see greater application of policies.
Here we focus on the 5 most popular policies as they account for 92.3\% of the instances, 73.6\% of users and 88.8\% of the posts. For completeness, we include the sum of the other less popular policies too. 
Figure~\ref{fig:policy time} presents the percentage of instances that activate each policy over time.
Across our measurement period, we observe a growth of 40\% in the total number of policies used.
This suggests that the use of policies is becoming more common.  
28.5\% of these policies are introduced by new instances coming online, with newly installed default policies, \eg 
\texttt{ObjectAgepolicy}, \texttt{TagPolicy} and \texttt{HashtagPolicy}.
The remainder are instantiated by pre-existing instance administrators that update their policies, suggesting a relatively active subset of administrators.

We also inspect the growth on individual instances. 
Overall, 42\% of instances add policies during our measurement period. 
Of these instances, 52.3\% enable only one extra policy and we see only a small minority (1.9\%) enabling in excess of 5 new policies (\eg \texttt{chaos.is} enables 13 and \texttt{poa.st} 12). A closer look at these instances show they mostly add common policies. 
However, we also see a wide range of other less common policies (\eg \texttt{KeywordPolicy}).

In contrast, the use of the \texttt{SimplePolicy}, with the most flexible range of moderation actions, has remained relatively stable.
Actions under the \texttt{SimplePolicy} have instance-wide effect and can effectively control instance federation. 
Overall, we only see 28.8\% of instances enabling this policy, without much growth across the measurement period (as seen in Figure~\ref{fig:policy time}). This could imply that administrators are unaware of this policy, do not have time to moderate their instances at this level or maybe find this policy too blunt (not fine-grained enough). The latter could lead to other issues, which administrators seek to avoid (\eg collateral damage~\cite{conext}).
It is also worth noting that the \texttt{SimplePolicy} is one of the most complex, and administrators potentially shy away from these more labour-intensive policies.
We argue that the diversity of policies could potentially overwhelm (volunteer) instance administrators (see Section~\ref{section:admin}).  This suggest that they require further support to automate this process (see Section~\ref{sec:watchlist}).

\section{Characterising Administrators}
\label{section:admin}

\subsection{Distribution of Administrators}
\label{subsection:admin_dist}

\begin{figure}[t]
 \centering
  \includegraphics[width=\linewidth]{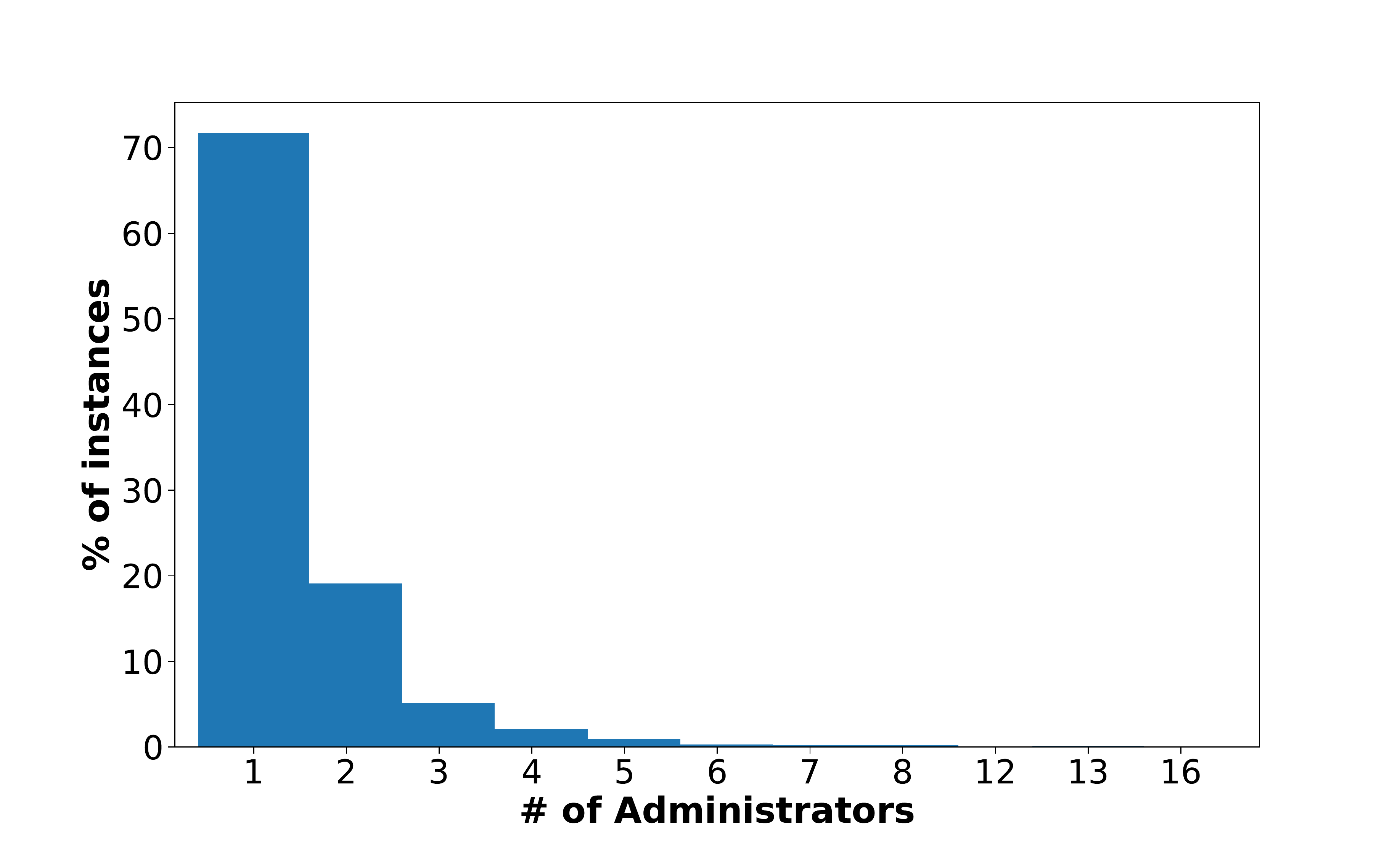}
  \caption{Instances (\%) by number of administrators.}
  \label{fig:inst_admin}
\end{figure}

\pb{Number of Administrators Per-Instance.}
We observe a total of 2,111 unique administrators from 1,633 instances (93.8\% of 1.74k).\footnote{The remaining instances do not publish their administrator(s) information.}
Figure~\ref{fig:inst_admin} presents the distribution of the number of administrators per instance.
Although a majority of instances (71.6\%) are managed by a single administrator,
we also see some instances with a larger number of administrators (\eg \texttt{rakket.app}: 16 and \texttt{poa.st}: 13).

\begin{figure}[t]
 \centering
  \includegraphics[width=\linewidth]{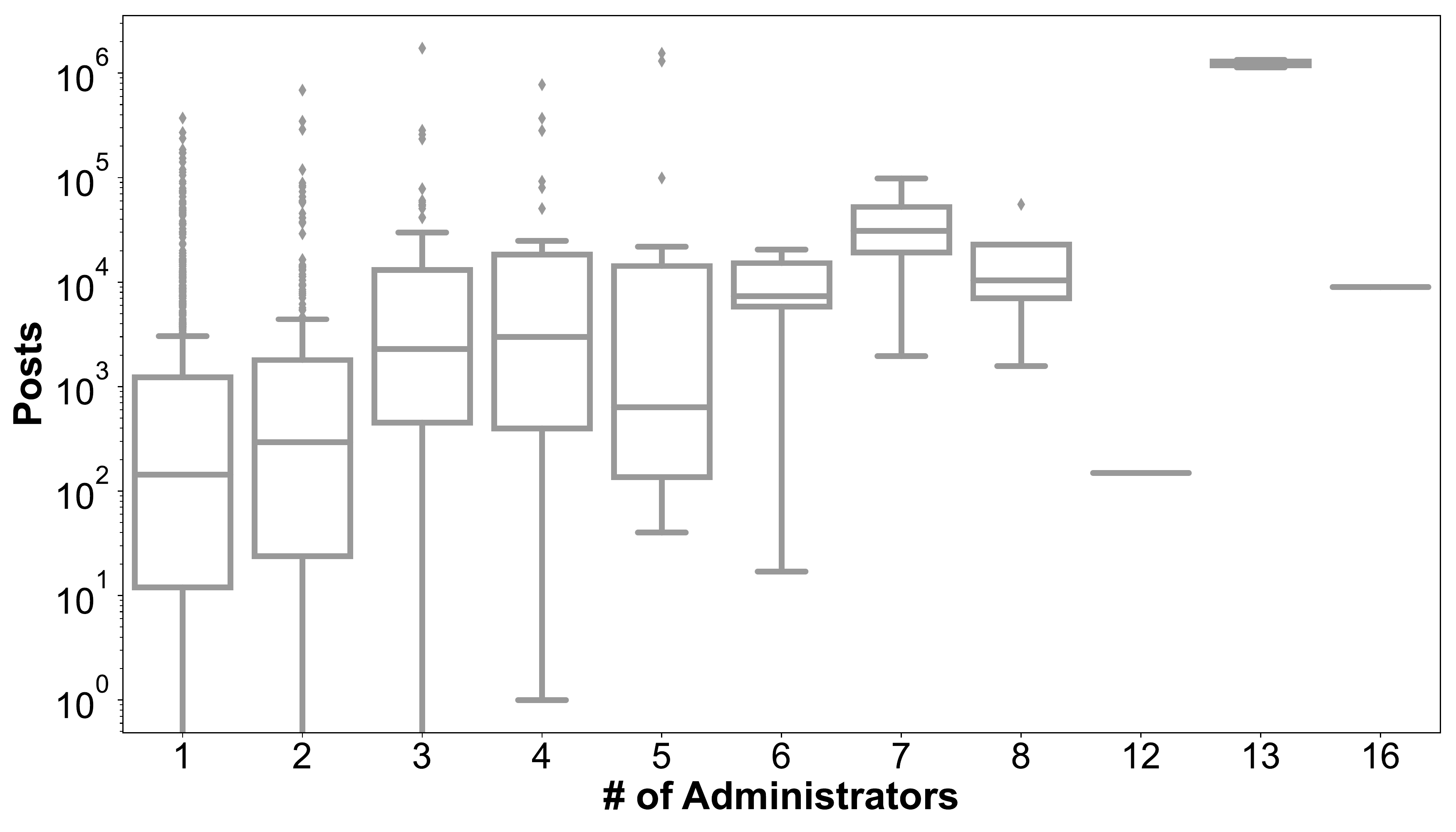}
  \caption{Box plot of the number of posts per instances with different number of administrators.}
  \label{fig:admin_number}
\end{figure}

\pb{Administrator Workload.}
We next test if the number of administrators increases proportionately to the number of posts.
We treat this as a rudimentary proxy for how much moderation must take place on an instance.
Figure~\ref{fig:admin_number} presents the distribution of posts on instances \vs the number of administrators.
Generally, we find that instances with more posts do have more administrators on average, \eg instances with multiple administrators have more posts, with a ratio of 6:1. However, this is driven by a few instances (\eg \texttt{poa.st}).

\begin{figure}[t]
 \centering
  \includegraphics[width=\linewidth]{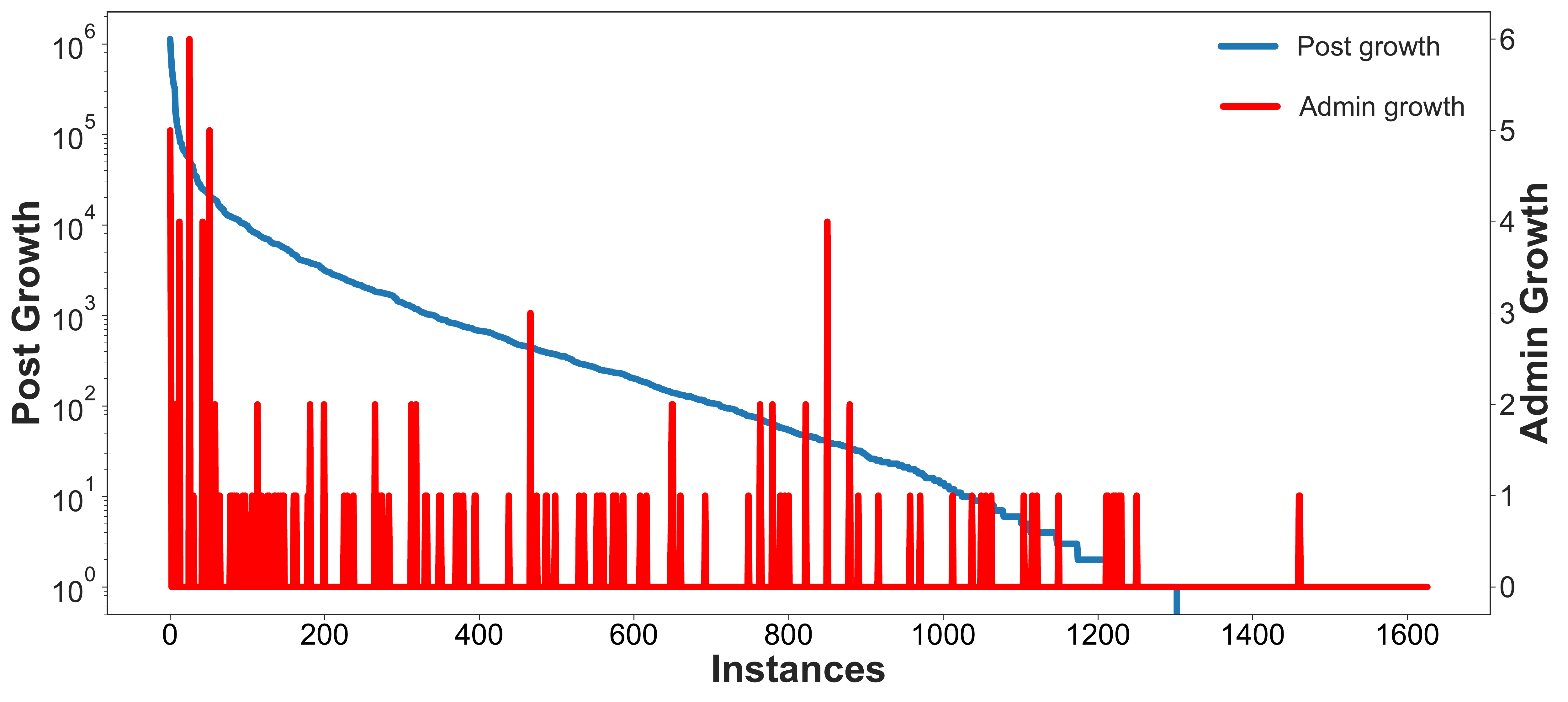}
  \caption{Per instance growth in the number of administrators (Y2-axis) and posts (Y1-axis). Individual instances are on the X-axis, sorted by the number of posts.} 
  \label{fig:adm_posts}
\end{figure}

Table~\ref{table:top10_most_growth} summarizes the top 10 instances that see the largest growth in administrators. Many of them are small instances with under 1000 users, and a proportionately small number of posts.
This suggests that administrator growth does not necessarily occur on the instances that need it the most.
To test if the number of administrators grow proportionately to the number of posts, Figure~\ref{fig:adm_posts} plots the \emph{growth} of administrators \vs the growth of posts on each individual instance during our data collection period. 
We see that a growth in posts on a given instance does not necessarily correspond to the recruitment of new administrators. In fact, only 6.9\% of instances record a growth in administrators during this period.
Overall, there is a weak correlation (Spearman coefficient of 0.19 for the number of posts \vs number of administrators). 
In total, we see a 60.3\% increase in the number of posts, but just a 35.6\% growth in administrators.
Unsurprisingly, instances that grow their administrator pool \emph{do} become more active. 
On average, instances with a growing number of administrators have 1.5x more policies than other instances. Specifically, looking at the policy with the most impact (\texttt{reject}), these instances apply it 1.8x more than others.
Interestingly, instances with an increasing number of administrators also have 4x more policies applied against them.

\subsection{Administrators' Response Lag}
\label{subsection:res}

The previous section has shown that administrators face a growing moderation workload.
To study this workload, we now look at how long it takes administrators to apply polices against particular instances. We focus on the \texttt{SimplePolicy} as this is clearly geared towards moderation, has instance-wide targeting, and lists the target instances.
For each \texttt{SimplePolicy} against a given instance, we compute the lag between the date of the implementation of the policy and the date when the targeted instance was first federated with.
This is a rudimentary proxy for how long it took an administrator to identify the problem. We temper our analysis with the fact that there could be many reasons for this delay, which we have limited vantage on.

\begin{figure}[t]
 \centering
  \includegraphics[width=\linewidth]{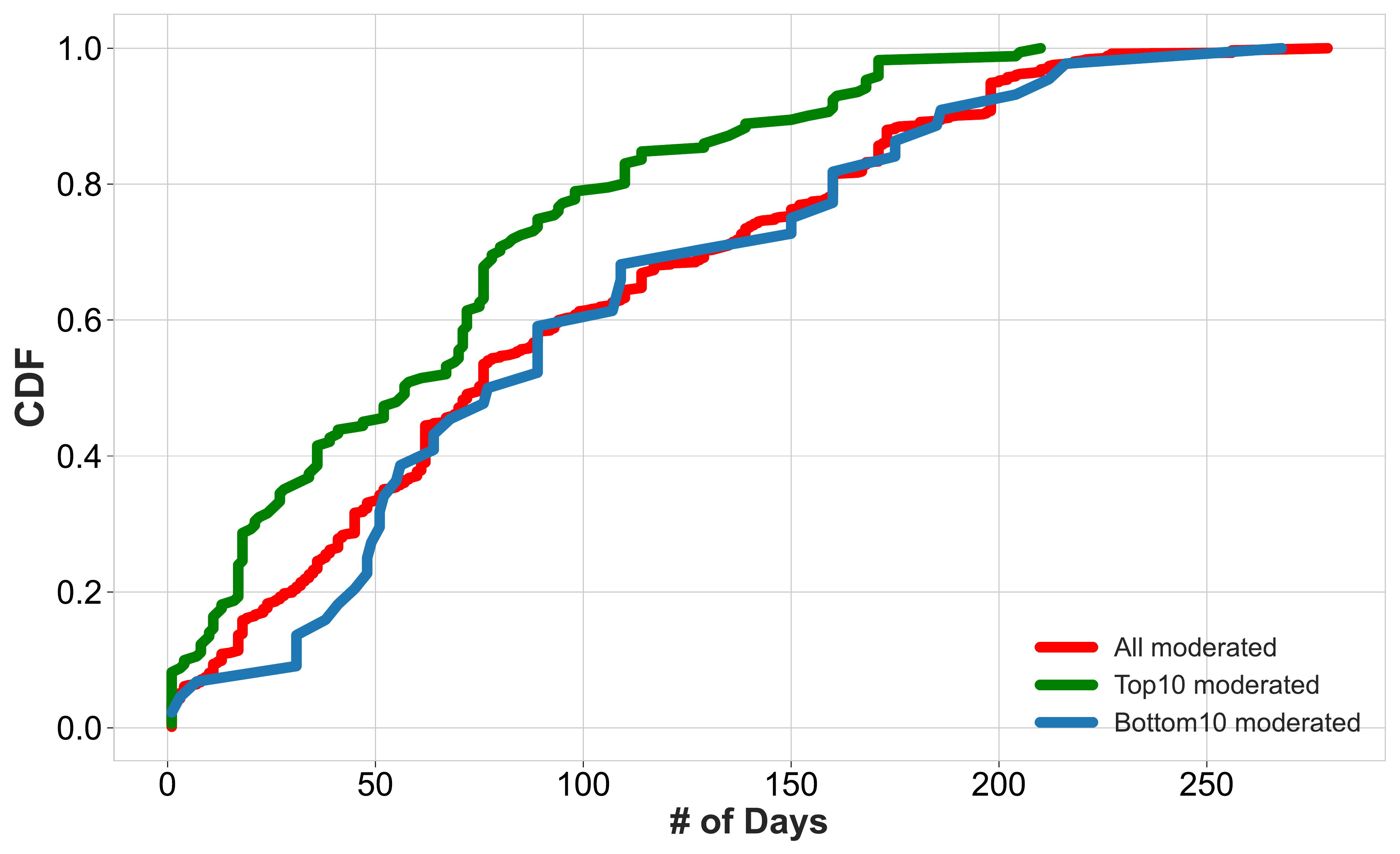}
  \caption{CDF showing the distribution of days from federation to moderation for all moderated instances. We also show results for the top 10 and bottom 10 instances, based on the number of policies applied against them.}
  \label{fig:fed_mod_cdf}
\end{figure}

\pb{Policy Creation Delay.}
Figure~\ref{fig:fed_mod_cdf} presents the distribution of delays (as defined above). 
Note, we exclude the 55\% of federations that occurred before the beginning of our data collection (as we cannot know their timestamp).
We plot the delay distributions for applying policies against: 
\one~All instances;
\two~``Controversial'' instances with the most policies applied against them (top 10); 
and
\three~``Benign'' instances with the fewest policies against them (bottom 10). 

It takes administrators an average of 82.3 days to apply any form of policy against other instances. Although, on average, it takes more time for a policy to be applied on the ``bottom 10'' instances than the "top 10" instances (74.7 and 59.5 days respectively), we see that there is a noticeable lag (almost 3 months) between federation occurring and policies being imposed.
This may suggest that administrators find it difficult to keep-up with the need to rapidly identify instances that justify policy imposition.

\begin{figure}[t]
 \centering
  \includegraphics[width=\linewidth]{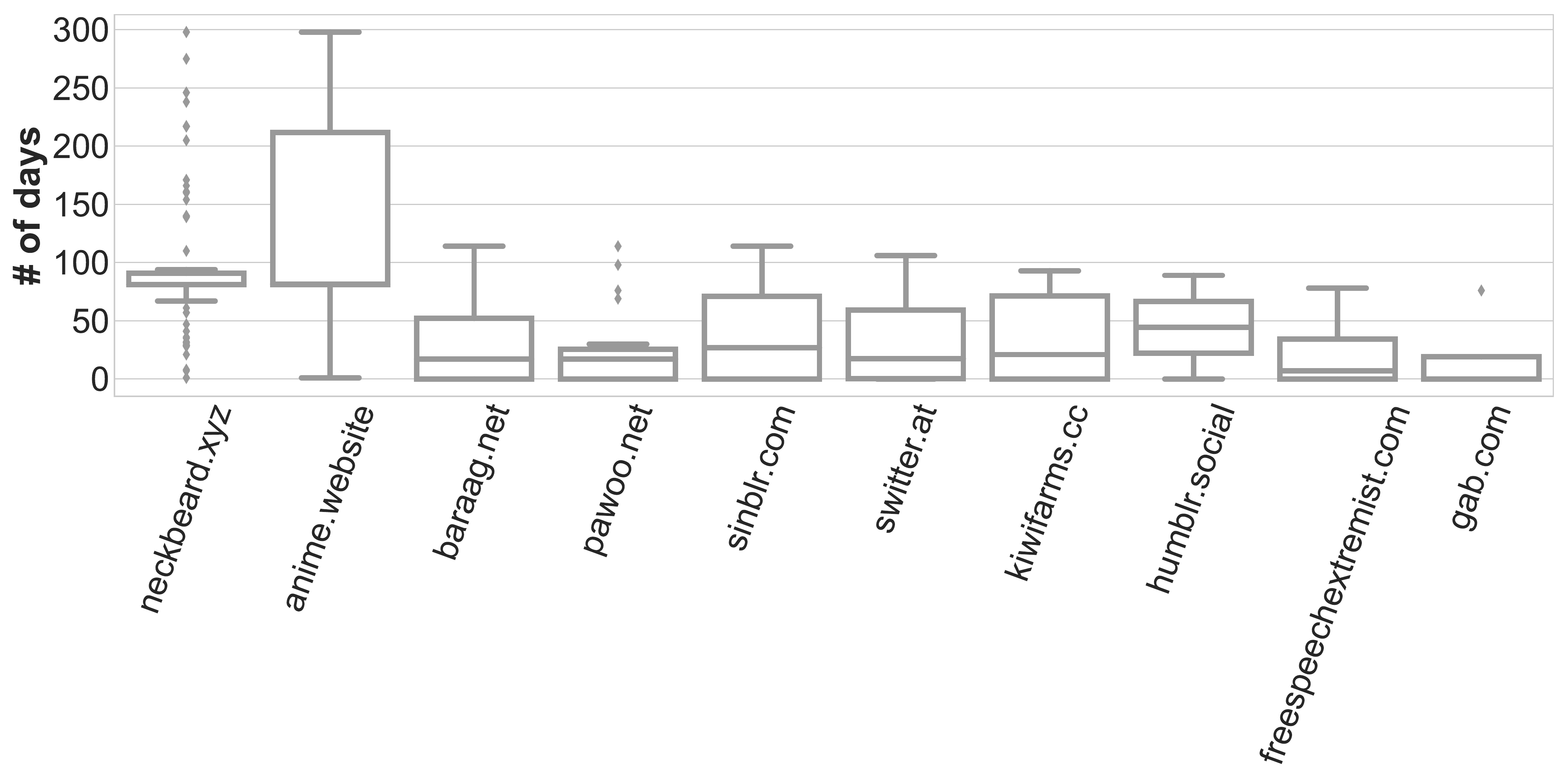}
  \caption{Box plot showing the distribution of the number of days from federation to the imposition of policies for the top 10 instances with the most policies applied against them.}
  \label{fig:fed_mod}
\end{figure}

\pb{Delay for Controversial Instances.}
We next extract the top 10 instances that receive the most policies targeted against them. For each one, Figure~\ref{fig:fed_mod} plots the distribution of delays (\ie how long it takes other instances to impose a policy against them).
In-line with expectations, we see that administrators take less time to apply policies against instances like \texttt{gab.com}, known for its right-wing stance (average of 19 days). 
However, we see much longer delays for other controversial instances that are less well-known (\eg \texttt{neckbeard.xyz}), averaging up to 98.4 days.
These instances are quite active, with significant growth in posts during our measurement period (\eg 
\texttt{neckbeard.xyz}: 789.4k and \texttt{kiwifarms.cc}: 469.2k). 
With other instances such as \texttt{anime.website} posting ``lolicon'' (suggestive art depicting prepubescent females), it is expected that policies would be swift, however, we see a very wide breadth of delays. 
The diverse nature of these administrator reactions indicates that any future automated moderation tools should be specialized to the  preferences of individual administrators.

\subsection{Administrators \& Moderators}
\label{subsection:mod}

\pb{Moderation Delegation.}
As administrators are responsible for a wide range of activities, they can delegate the task of content moderation to select individuals. These accounts are referred to as \textbf{moderators}. Of our 1.74k instances, 47\% of them (819) expose information in our dataset.
From these, only 12\% (98) of instances have assigned the role of moderator to any other accounts. Of these, 73.5\% (72) of the instances have the administrator also doubling as a moderator, while 29.6\% (29) of the instances assign the entire moderator role to an account that is not the administrator. This implies that only 3.5\% of instances have dedicated account(s) assigned the role of moderator.

\begin{figure}[t]
 \centering
  \includegraphics[width=\linewidth]{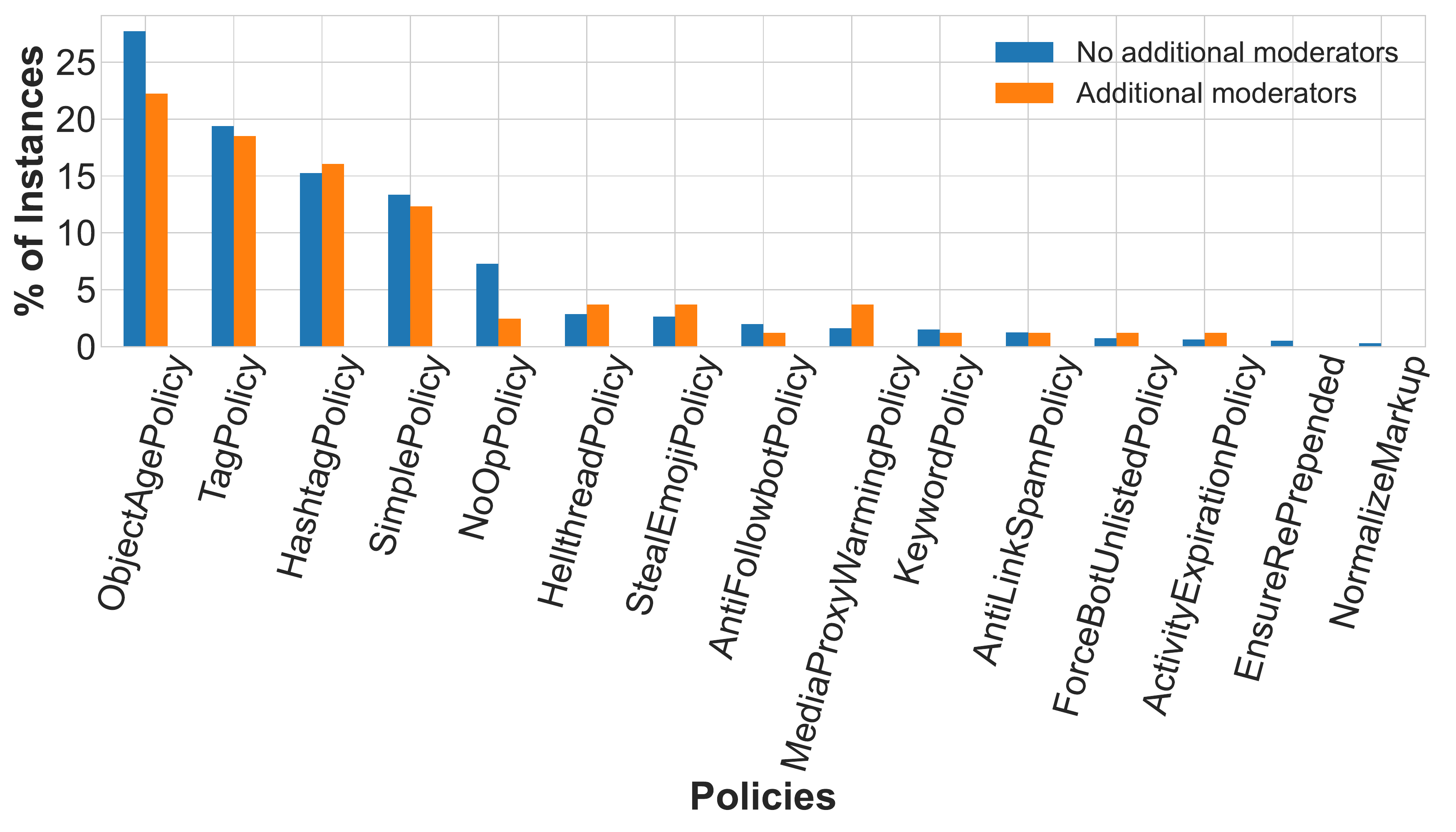}
  \caption{The percentage of instances that enable the top 15 most popular policies. We separate instances into two groups: \one~Instances without additional moderators; and \two~instances with additional moderators outside of the administrator set.}
 
  \label{fig:mod_policy}
\end{figure}

\pb{Are moderators helpful?}
We conjecture that instances with dedicated moderators outside of their administrator team might be swifter in the application of policies.
Figure~\ref{fig:mod_policy} shows the percentage of instances that enable the 15 most popular policies (Figure~\ref{fig:policies}).
We present two bars for each policy:
\one~Instances with additional moderators (who are not an administrator);
and
\two~Instances without additional moderators. 
There is a broadly similar distribution across these two groups. 
However, we notice that instances without additional moderators have approximately 3x more of the \texttt{NoOpPolicy} configured. Recall, this is the default state of an instance and allows any content to be imported. This begins to suggest that instances with additional moderators do pay greater attention to policies.

\begin{figure}[t]
 \centering
  \includegraphics[width=\linewidth]{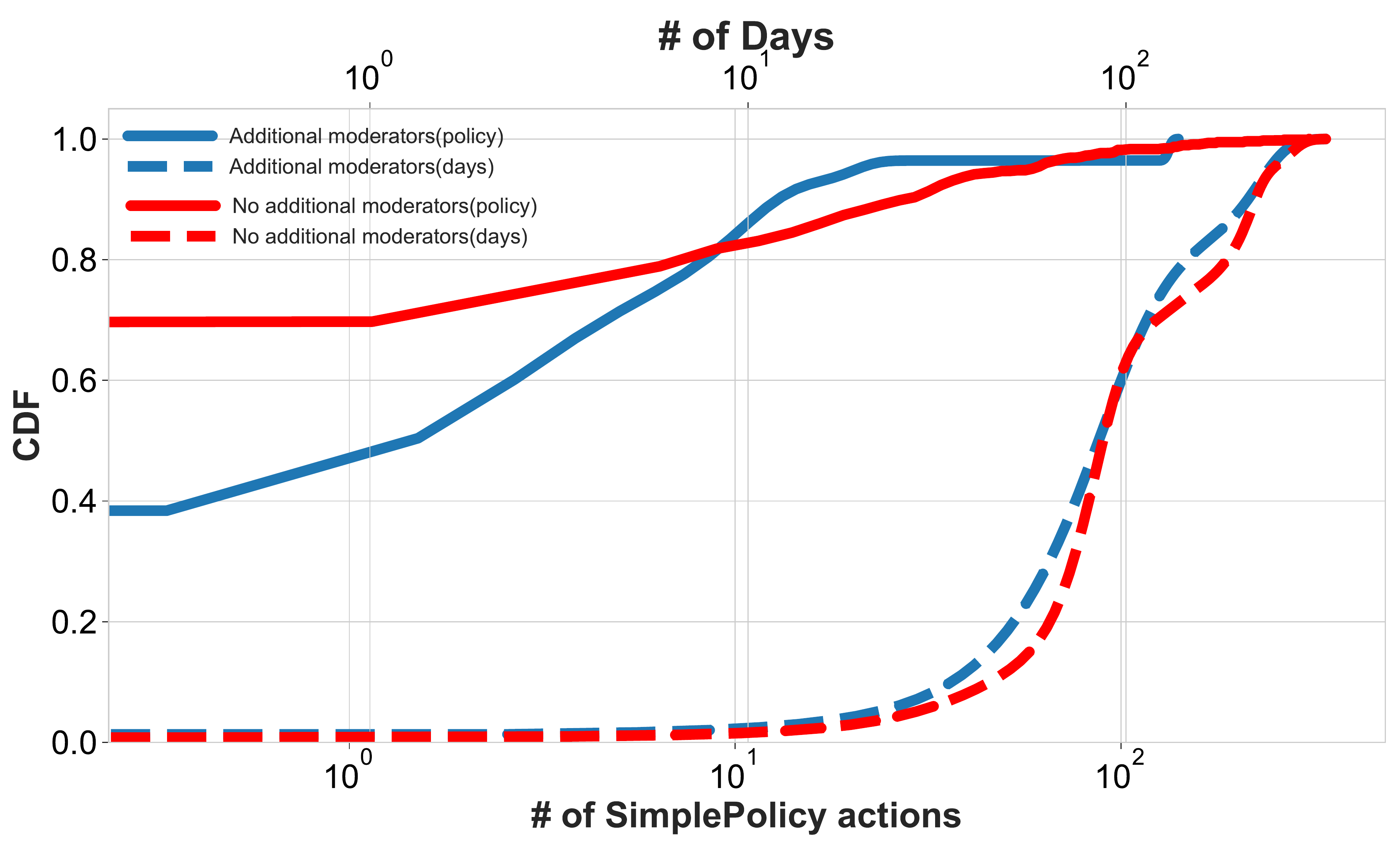}
  \caption{CDF of the number of \texttt{SimplePolicy} actions per instance (X1-axis) and
  the lag (in days) for instances to impose a policy after federation (X2-axis). 
  We separate instances into \one~those with dedicated moderators; and \two~those without dedicated moderators. 
  }
  \label{fig:simpol_mod}
\end{figure}

We expand this analysis in  Figure~\ref{fig:simpol_mod}, where we show the number of \texttt{SimplePolicy} actions and the delay to apply a policy after federation (in days) for instances in the two groups. We use the \texttt{SimplePolicy} for this analysis as it is the only moderation policy with instance-wide targeting and a list of targeted instance domains. 
The plot shows that instances with moderators take less time (average 103 days) to impose a \texttt{SimplePolicy} after federation, compared to instances without dedicated moderators (average 111 days).
The figure also shows a marked difference in the number of instances that apply the \texttt{SimplePolicy}.
Only 38\% of the instances with dedicated moderators apply no \texttt{SimplePolicy} actions, compared to 70\% for those without. 
This confirms that instances with additional moderators are more proactively moderated.

\section{WatchGen: Automating Moderation}
\label{sec:watchlist}

Our results indicate that moderation is labor-intensive. We now explore techniques to assist administrators. 
We propose \emph{WatchGen},\footnote{\url{https://github.com/anaobi/WatchGen.git}} a tool that recommends  to administrators a ``watchlist'' of  instances that may require federated moderation. This watchlist must be on a per-instance basis, as different administrators may have varying views on what is considered appropriate for the instance they manage. 
WatchGen, helps administrators to more proactively  identify instances requiring  attention with regards to content moderation. 
We build WatchGen by compiling a large feature set for each instance, and experimenting with a number of classification models to flag instances that are more likely to require attention.

\pb{Feature Selection.}
We first extract features for each instance.
These features include information about user  (\eg number of users) and administrator activities with respect to moderation (\eg number of rejected instances). We also extract features from post content (\eg number of hate words in posts). 
We experiment with a total of 38 features (see Table~\ref{table:features}).
Through extensive manual experimentation, we distil this down to the 16 most determinant features (highlighted in Table \ref{table:features}).

\pb{Model Training.}
Next, we train multiple machine learning models using the \texttt{sklearn} library, and GridSearchCV within 5-fold cross-validation to find the optimal hyper-parameter settings.
We detail below the hyperparameters for each model.

\pb{Logistic Regression (LR).}
We only tune the C hyperparameter. This regularization parameter controls how closely the model fits to the training data. We test for best value of "C" using the values \{0.001, 0.01, 0.1, 1, 10, 100, 1000\}.

\pb{Multilayer Perceptron (MLP).}
We tune three hyperparameters: \one~hidden layer-size: dictates the number of hidden layers and nodes in each layer. We use a single hidden layer with varying hidden layer-sizes \{10, 50, 100\};
\two~activation function: determines the type of non-linearity introduced into the model. We employ 3 activation functions \{relu, tanh, logistic\};
and
\three~learning rate: we tune how the initial learning rate parameter changes in finding the optimal model using \{constant, invscaling, adaptive\}. 

\pb{Random Forest (RF).}
We tune 2 hyperparameters. \one~n\_estimators: the number of independent trees (estimators). We test using 3 values \{5, 50, 250\};
and
\two~max\_depth: the depth of the trees. We test for best result using 6 different depths \{2, 4, 8, 16, 32, None\}.

\pb{Gradient Boosted Trees (GB).}
We tune three hyperparameters.
\one~n\_estimators: The number of independent trees (estimators). We test with 4 value \{5, 50, 250, 500\}; \two~max\_depth: The depth of the trees. We test with 5 values \{1, 3, 5, 7, 9\}. \three~Learning rate: This impacts the speed and granularity of the model training. We test  5 values \{0.01, 0.1, 1, 10, 100\}.

\subsection{Generating a Global Watchlist}
\label{subsection:predicting}

\pb{Task.}
We first assume a WatchGen central broker that compiles a global pool of training data, collected from all instances through their public APIs (similar to us in Section~\ref{section:data}).
We use this global pool of training data, with an 80:20 split, to predict if a given instance will be subject to \emph{any} policy (by any other instance).
We then produce a  `watchlist' of instances that may be worthy of attention. 

To investigate how long it would take to garner sufficient data to train WatchGen, we also train several models on datasets covering increasing time windows. 
We first train on one month of data and increase the training dataset by one month at a time (up to 9 months). 
For our test dataset, we use the data remaining after the training snapshot.

\begin{figure}[t]
  \includegraphics[width=1\columnwidth]{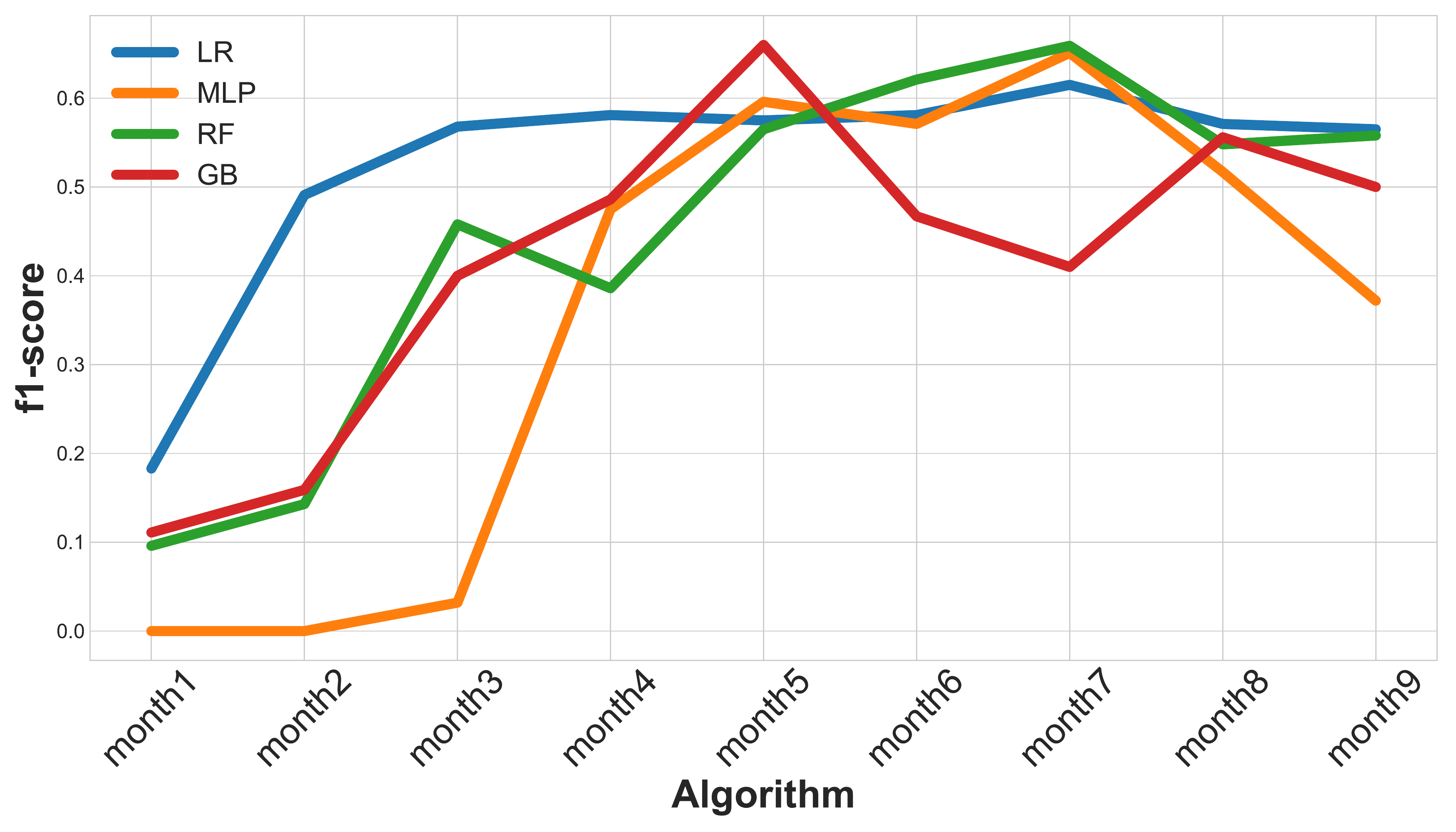}
  \caption{Time series  of  f1-scores for the Logistic Regression, Multi-Layer Perceptron, Random Forest and Gradient Boosted Trees models. Note that we exempt month 10 as this leaves insufficient test data.}
  \label{fig:f1_time}
\end{figure}

\pb{Results.}
Table \ref{table:model} summarizes the result with the global pool of training data (80:20 split) with Random Forest being the best performing model (f1=0.77).
Recall, that we also run experiments with a training set based on varying time windows.
Figure~\ref{fig:f1_time} presents the f1 scores based on the size (duration) of the training set.
We observe that it takes at least 5 months for a model to achieve its best score  (\eg Gradient Boosted Trees is month 5 and Random Forest in month 7). Note that the training sets are different from Table \ref{table:model} and hence the scores differ.  

\begin{figure*}
  \begin{subfigure}[b]{0.33\linewidth}
    \includegraphics[width=\linewidth]{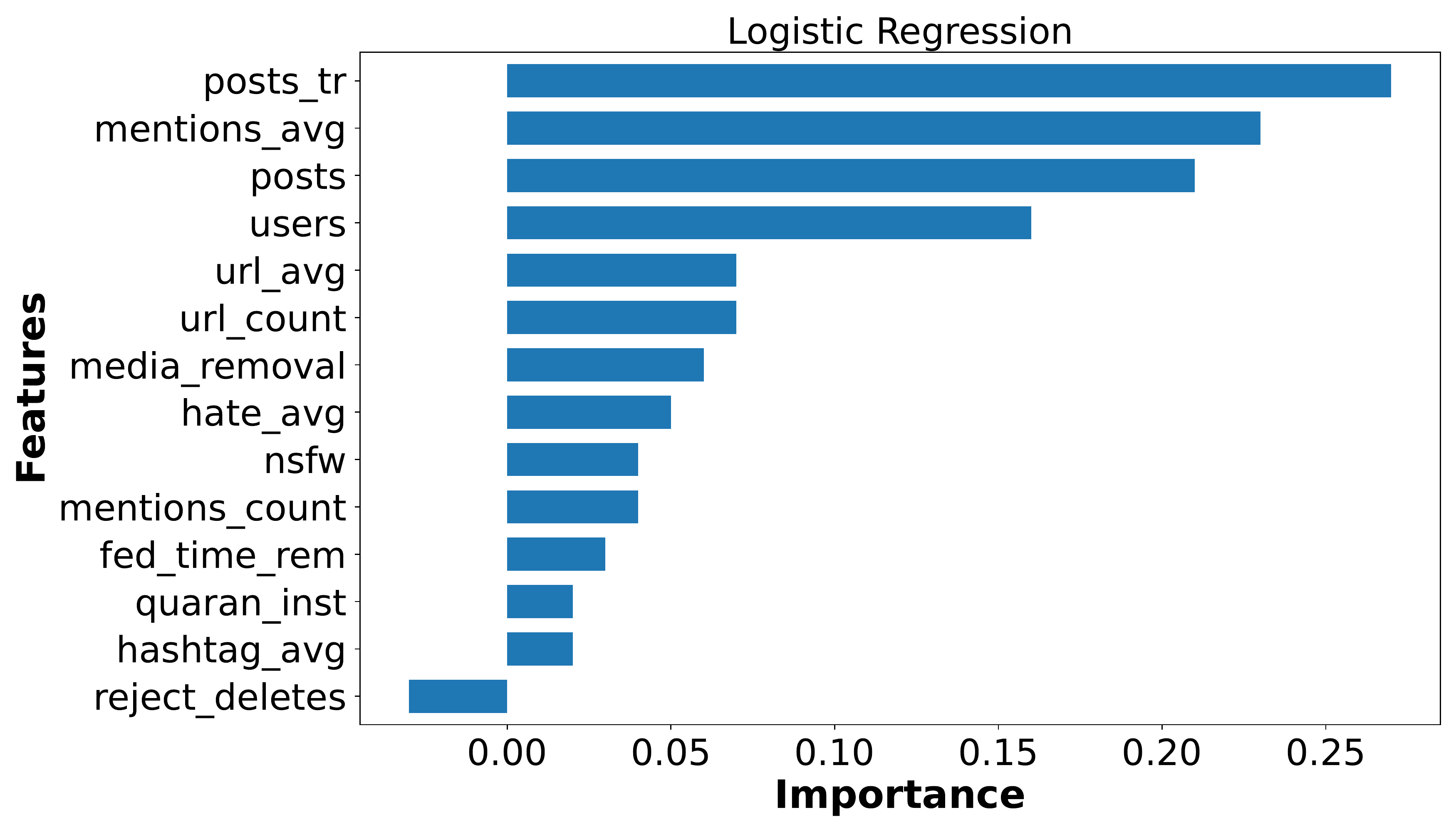}
    \caption{Logistic Regression}
    \label{fig:LR}
  \end{subfigure}
  \begin{subfigure}[b]{0.33\linewidth}
    \includegraphics[width=\linewidth]{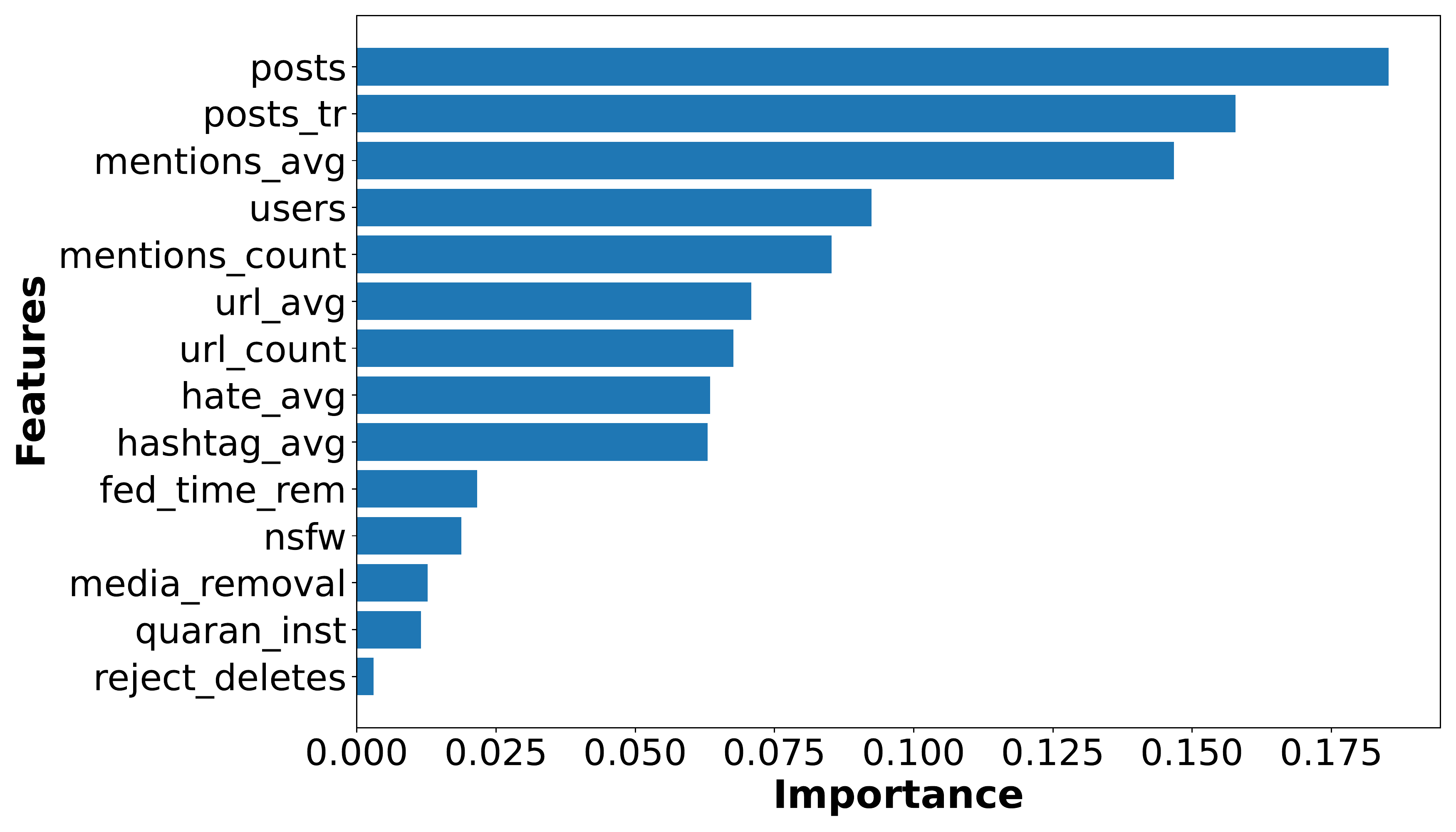}
    \caption{Random Forest}
    \label{fig:RF}
  \end{subfigure}
  \begin{subfigure}[b]{0.33\linewidth}
    \includegraphics[width=\linewidth]{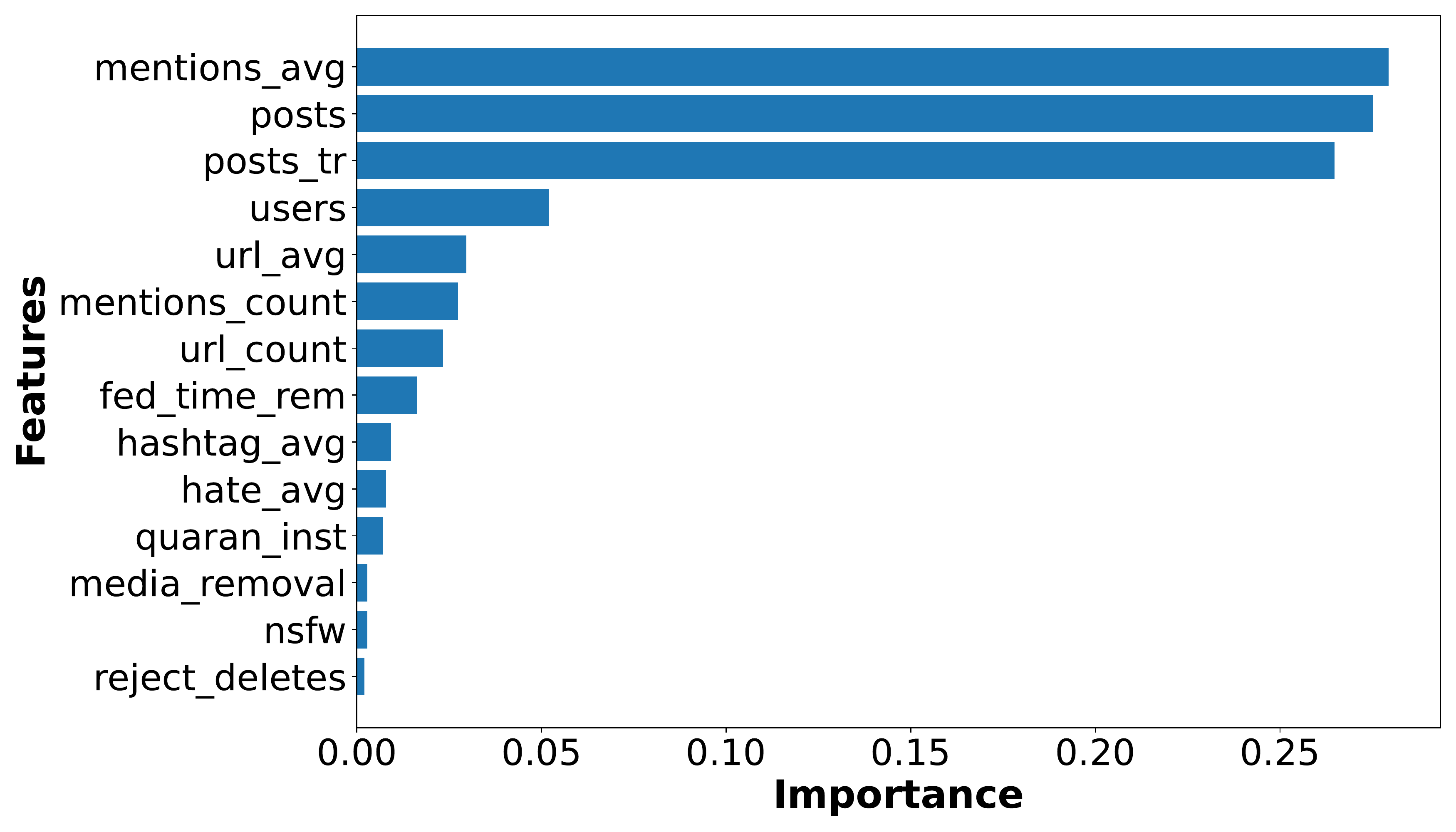}
    \caption{Gradient Boosted Trees}
    \label{fig:GB}
  \end{subfigure}
  \caption{Feature importance for our explainable models.}
  \label{fig:feat_imp}
\end{figure*}

\pb{Feature Importance.}
We next inspect which features are most important. This sheds insight into which characteristics are most related to triggering policies. We use the in-built functions for feature importance.
Figure~\ref{fig:feat_imp} presents the feature importance for the explainable models.
We see that the top 3 features (transformed post, average number of mentions in a post, and number of posts on an instance) are all related to the number of posts on an instance. This suggests that the likelihood of an instance having a policy applied against it is closely related to the amount of content its users post. In other words, the more users and posts on an instance, the higher the probability of having a policy applied against it. This is expected as such instances are likely to attract more attention.

Features such as the number of mentions and  hate words in the posts also play an important role. 
This is in-line with prior work that  observed how mentions and quote retweets result in more attention~\cite{garimella2016quote}.
To better understand the importance of these secondary metrics, we retrain the model without the two top features (number of posts and transformed posts). 
We show the results in Table~\ref{table:model2}. 
Confirming our prior assertion, we retain relatively good performance. For Random Forest, we attain an f1 of 0.62 (\vs 0.77 with the full feature set in Table~\ref{table:features}). This confirms that these other factors play an important role in determining if an instance has a policy applied against it. 
In other words, in addition to the size of an instance, other features are required to obtain a fairly good prediction of instances being subject to any policy.

\begin{table}

\begin{tabular}{llrll}
 \toprule
Algorithm & \begin{tabular}[c]{@{}r@{}}Acc.\end{tabular} & Prec. & Recall & f1 score \\
\midrule
 Logistic Regression &  0.86 & 0.85 & 0.34 & 0.49\\
 Multi-Layer Perceptron & 0.57 &  0.34 & 0.42 & 0.53 \\
 Random Forest & 0.92 &  0.88 & 0.68 & 0.77 \\
 Gradient Boosted Trees &  0.89 & 071 & 0.71 & 0.71 \\
                        
\bottomrule
\end{tabular}

\caption{WatchGen performance results when using global training pool and the full feature set.}

 \label{table:model}
\end{table}

\subsection{Generating a Local Watchlist}

\pb{Task.} 
Our prior WatchGen models assume a central pool of training data, aggregated from all instances. This may be infeasible in practice due to the decentralized nature of the Fediverse.
Hence, we next investigate how well our best model (Random Forest) performs when decentralizing the training process.
For each instance, we extract its federated peers and exclusively build a local training set from their data (using the features highlighted in Table \ref{table:features}).
For each pair of instances, we  tag whether or not a directed policy is imposed, \ie each instance only considers the policies it locally sees.
Finally, each instance trains its own local model using the first 8 months of data (and tests on the last 2).
This creates one independent model per-instance. 
Based on this, WatchGen predicts whether a policy will be applied against the instance.

\begin{figure}[!h]
  \includegraphics[width=0.85\columnwidth]{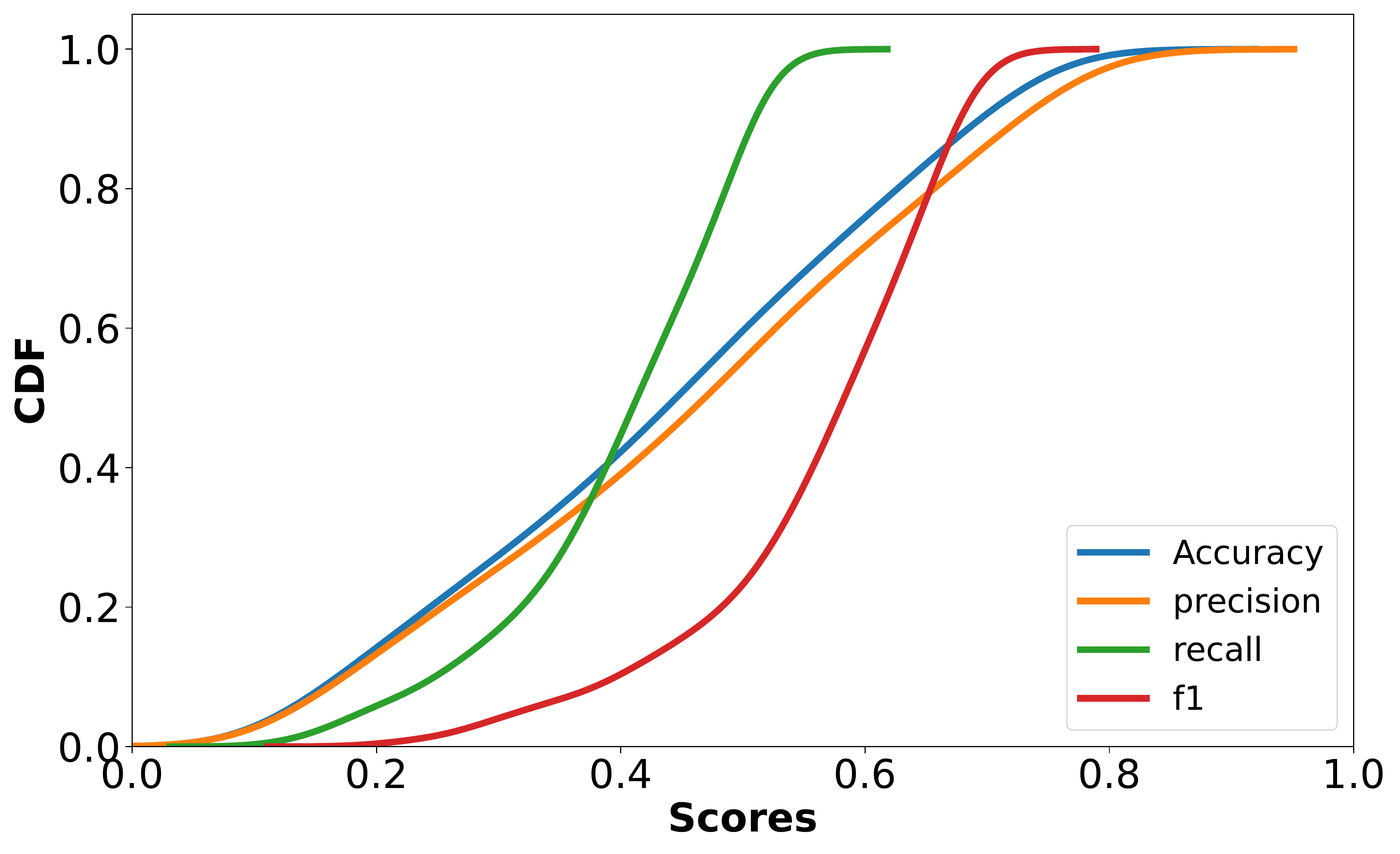}
  \caption{CDF of per-instance performance for Random Forest  trained on data from local and federated instances.}
  \label{fig:per_inst}
\end{figure}

\pb{Results.}
Figure~\ref{fig:per_inst} presents the distribution of performance metrics per-instance.
As expected, we observe an overall performance drop compared to the prior task based on a global model.
Instances attain an average f1 score of 0.55.
This is largely due to the significant reduction in per-instance training data.
That said, we observe a wide array of performances across the instances:
42.6\% of instances achieve above 0.6 f1, with a tail of 8.3\% attaining below 0.4 f1. We find that performance is impacted by the training set size.
Instances that perform relatively well ($>=$0.6 f1), tend to be larger (\ie more posts and users). For example, 65.4\% of the best performing instances ($>=$0.6 f1) have a local post count of over 50k (\eg \texttt{neckbeard.xyz} and
\texttt{freespeechextremist.com}). 
In contrast, only 4.4\% of instances that perform poorly ($<$0.6 f1) have over 50k posts (\eg \texttt{princess.cat} 
and
\texttt{sleepy.cafe}).
This implies that as instances grow, their local performance will improve.
The above experiments show that instances \emph{can} use these locally trained models to generate a personalized watchlist of instances they peer with. 
Thus, we argue that these automatically compiled lists can helps administrator pay attention to these instances.

\section{Related Work}
\label{sec:related}

\pb{Social Network Studies.}
Extensive work has been carried out in the area of online social networks. However, most of these are on centralized social networks (\eg Facebook and Twitter)~\cite{truad, cha, kwak, ravi, yong, seth}. A number of these look at the anatomy of social graphs~\cite{ugander} and moderation challenges~\cite{ibosiola2019watches}. Others look into areas ranging from the evolution of user activities to demographics~\cite{insta, bimal}.
In contrast to Pleroma, these social networking platforms tend to rely on central (commercial) administrators and moderators~\cite{wauters2014optimizing}.

\pb{Fediverse and Decentralized Web.}
Only a small set of studies have focused on the Fediverse or Decentralized Web applications. 
Raman \etal looked at the challenges in the Fediverse, with a particular focus on the infrastructure and resilience of Mastodon~\cite{Mastodon}.
Trautwein \etal studied the Inter Planetary File System (IPFS), a decentralized storage solution~\cite{trautwein2022design}.
Guidi \etal and Datta \etal studied the structure, data management, and privacy aspects of decentralized social networks~\cite{guidi, datta}. 
Recent works have examined the standardization of related protocols~\cite{khare2022web,mcquistin2021characterising}. 
Bielenberg \etal analyzed  the growth, topology and server reliability of Diaspora (a decentralized social network) \cite{bielenberg2012growth}. 
Similarly, Zignani \etal studied the evolution of the Mastodon social graph~\cite{Matteo}. 
Our work differs in that we focus on exploring \emph{administrator} actions within the Fediverse.

\pb{Online Moderation.}
Prior work has investigated the roles that volunteer moderators play in platforms like Twitch~\cite{twitch}.
Text-based content classification and filtering has been extensively studied too. These include computational techniques to detect cyberbullying~\cite{bullying, jun, ziems}, anti-social posting~\cite{pro-eating, antisocial, insult, zia2021racist}, and hate speech \cite{burnap, will, amir, zhi, iqbal2022exploring, zeerak, javed2020first}. 
These models have proven effective in reducing the workload of human moderators. For example, Cheng et. al. \cite{antisocial} use random forest and logistic regression classifiers to predict whether a user will be banned, reducing the manual load on moderators. Similarly, Zia \etal~\cite{zia2022toxicity} look at detecting the spread of toxic posts specifically in Pleroma (although not administrator reactions). 
In our prior work, we also studied the use of federation policies~\cite{conext}.
Here, we build on this, with a focus on the actions undertaken by administrators. We further propose WatchGen to assist administrators. 
To the best of our knowledge, this is the first large-scale study of administrator activities in the Fediverse. We hope that this can further contribute to the wider understanding of moderation in other platforms.

\section{Conclusion and Discussion}
\label{sec:conclusion}

We have studied instance administrators in a popular Fediverse platform, Pleroma. 
Although  66.9\% of instances are still running on default policies, we observe an uptake of more sophisticated management functions.
We find evidence that some administrators may become overwhelmed with the growing number of posts and users they must manage.
For instance, it takes an average of 82.3 days for administrators to apply any policy against a newly federated instance.
Another sign of the overhead is that just 3.5\% of instances share the load across multiple moderators. 
This lack of moderators may come with challenges: instances with fewer moderators tend to employ less sophisticated policy strategies (\eg 70\% of them apply no \texttt{SimplePolicy} actions).
To alleviate this, we have proposed WatchGen, a tool that identifies instances in need of closer attention. We show that WatchGen can  predict which instances will later have a policy imposed (f1 = 0.77). 

Our study opens up a number of lines of future work. First, we wish to expand our work to cover other Fediverse platforms, \eg Mastodon or PeerTube. Second, we plan to experiment with alternate feature sets that can better identify instances that will later require policy attention. Through this we hope to improve WatchGen and pilot its deployment. Last, we want to perform a qualitative study to better understand the subjective opinions of administrators that underlie these trends. We conjecture that such qualitative insights might be invaluable for improving WatchGen.

\section*{Acknowledgements}
This research was supported by EPSRC grants EP/S033564/1, EP/W032473/1, UKRI DSNmod (REPHRAIN EP/V011189/1), and EU Horizon Framework grant agreement 101093006 (TaRDIS).\

\small
\bibliographystyle{abbrv} 
\bibliography{pleroma_arxiv}

\appendix

\section{Appendix}

\begin{table*}[]
\resizebox{2.1\columnwidth}{!}{
\begin{tabular}{llrrrrrrrl}
 \toprule
Policy & \begin{tabular}[c]{@{}r@{}}Description\end{tabular} & \% Instances& \% Users & \% Posts &  Growth in Inst. & \% Growth in Inst \\
\midrule
                                                                  \\
ObjectAgePolicy                & Applies action based on post age     & 74.80  & 57.00  &   65.30 & 352 & 73.50\%\\
TagPolicy                & Applies policies to individual users based on tags   &  58.50 & 39.40 &  31.30 & 509 & 707.40\%\\
HashtagPolicy           & List of hashtags to apply actions against      & 36.40  & 16.20  &  21.20 & 479 & 15,833.00\%\\
SimplePolicy            & Wide range of actions applied against instances   & 28.80 & 39.70 &  36.30 & 83 & 30.70\%\\
NoOpPolicy       & Default state of an instance      & 11.50 & 5.90 &  3.70 & -98 & -63.70\%\\
StealEmojiPolicy       & List of hosts to steal emojis from   & 7.00 & 6.10 &  5.40 & 29 & 80.50\%\\

HellthreadPolicy          & Performs action when a threshold of mentions is reached      & 6.50  & 10.90 &  19.80 & 21 & 42.80\%\\
AntiFollowbotPolicy            & Stops bots from following users on the instance   &  4.50  & 6.20 &  6.90 & 13 & 40.60\%\\
MediaProxyWarmingPolicy       & Crawls attachments using their MediaProxy URLs     & 3.60  & 7.00 &  8.30 & 16 & 72.70\% \\
KeywordPolicy       & Matches a pattern in a post for an action to be taken    & 23.00  & 19.40 &  10.00 & 9 & 36.00\% \\

ForceBotUnlistedPolicy          & Makes all bot posts to disappear from public timelines     & 2.70    & 7.00 &  5.50 & 27 & 675.00\%\\
AntiLinkSpamPolicy   & Rejects posts from likely spambots by rejecting posts from new users that contain links       &  2.70  & 6.70 &  6.80 & 12 & 85.70\%\\
ActivityExpirationPolicy       & Sets a default expiration on all posts made by users of the local instance.  & 1.30  & 1.20 &  0.73 & 11 & 366.60\%\\
EnsureRePrepended       & Rewrites posts to ensure that replies to posts with subjects do not have an identical subject   & 1.30 & 0.40 &  1.80 & 6 & 66.60\%\\

NormalizeMarkup      & processes messages through an alternate pipeline                                      & 0.9 & 4.2 &  1.4 & 6 & 150\%\\

\bottomrule
\end{tabular}}

\caption{The top 15 policies applied by administrators with the percentage of instances applying the policies. It shows the percentage of users and posts on the instances applying them, and their growth during our measurement period.}

 \label{table:pol}
\end{table*}

\begin{table*}[]
\resizebox{1.\linewidth}{!}{
\begin{tabular}{lrrrrrrrrrrrrrrr}
 \toprule
Instances & \begin{tabular}[c]{@{}r@{}} Admin \\ growth  \end{tabular} & \# Admins & Users &\begin{tabular}[c]{@{}l@{}}User\\ Growth\end{tabular} & Posts & \begin{tabular}[c]{@{}l@{}}Post\\ Growth\end{tabular} & \begin{tabular}[c]{@{}l@{}}Hate\\ count\end{tabular} & \begin{tabular}[c]{@{}l@{}}URL\\ count\end{tabular} & \begin{tabular}[c]{@{}l@{}}Mentions\\ Count\end{tabular} & \begin{tabular}[c]{@{}l@{}}Hashtag\\ Count\end{tabular} & nsfw & \begin{tabular}[c]{@{}l@{}}Media\\ Removal\end{tabular} & \begin{tabular}[c]{@{}l@{}}Federated\\ Timeline\\Removal\end{tabular} & Reject & \begin{tabular}[c]{@{}l@{}}Quaran\\-tined \end{tabular}\\
\midrule

disqordia.space               & 6 & 8    & 53 & 33  & 55.5k & 5.1k & NA & NA & NA & NA & 3 & 2 & 15 & 71 & 17      \\

poa.st           & 5 & 13                            & 9.7k  & 9.46k & 1.14m & 453.3k & 78.2k & 19.5k & 60.8k & 20.1k & 4 & 3 & 2 & 1 & 0\\

pleroma.nobodyhasthe.biz              & 5  & 6    &  128 & 79 & 20.5k & 1.12k &42.8k & 2.6k & 42.2k & 2.2k & 0 & 1 & 0 & 2 & 0 \\

pleroma.pt            & 4 & 7     &  450 & 448 & 24.9k & 1.8k & 449 & 75 & 246 & 29 & 8 & 5 & 0 & 1 \\

pleroma.foxarmy.ml       & 4 & 5      & 8 & 7 & 40  & 7 & NA & NA & NA & NA & 0 & 0 & 0 & 0 & 0                        \\

varishangout.net         & 4 & 7      & 924 & 856  & 98.5k & 2.6k  & 4 & 1 & 0.0 & 3 & 0 & 3 & 9 & 6 & 0           \\
mindset.rage.lol          & 3 & 5        &  10 & 8  & 635 & 444 & NA & NA & NA & NA & 0 & 0 & 0 & 0 & 0                  \\
neckbeard.xyz       & 3 & 13              & 2k & 1.22k & 1.34m &789.4k & 883 & 136 & 607 & 177 & 0 & 0 & 0 & 2 & 0              \\
fedi.absturztau.be      & 2 & 4          & 900 & 463 & 775.8k  & 327k  & 12.9k & 1.9k & 9.5k & 2.4k & 3 & 0 & 0 & 14 & 12           \\

childpawn.shop       & 2 & 3& 183 & 176 & 3.8k  & 441 & NA & NA & NA & NA & 0 & 0 & 0 & 0 & 0                   \\

\bottomrule
\end{tabular}}

\caption{Top 10 Instances with the largest increase in number of administrators during our measurement period.}

 \label{table:top10_most_growth}
\end{table*}

\begin{table*}
\begin{tabular}{llrll}
 \toprule
Algorithm & \begin{tabular}[c]{@{}r@{}}Acc.\end{tabular} & Prec. & Recall & F1 score \\
\midrule
 Logistic Regression &  0.73  &    0.24  & 0.20 & 0.21\\
 Multi-Layer Perceptron & 0.81 &  0.00 & 0.05 & 0.10 \\
 Random Forest & 0.87 &  0.69 & 0.57 & 0.62 \\
 Gradient Boosted Trees &  0.87 & 0.73 & 0.54 & 0.62 \\
                        
\bottomrule
\end{tabular}

\caption{WatchGen performance results using global training pool and excluding post features (number of posts and transformed posts).} 

 \label{table:model2}
\end{table*}

\begin{table*}[]
\resizebox{2\columnwidth}{!}{
\begin{tabular}{lrrrrrrl}
 \toprule
Feature & \begin{tabular}[c]{@{}r@{}}\#Description\end{tabular} & \#Representation & Number  \\
\midrule
\rowcolor{Gray}
Users                & Number of users registered on an instance                                                                  & Count  & 133.8k    \\
\rowcolor{Gray}
posts                & Number of posts by users on an instance                                                                  &  Count  & 29.9m \\
\rowcolor{Gray}
hate\_count           & Number of hate words on an instance from \url{hatebase.org}                             & Count  & 36m   \\
\rowcolor{Gray}
url\_count            & Number of URLs in user posts on an instance                                            &  Count  & 4.8m \\
\rowcolor{Gray}
reject       & Number of instances to completely reject any flow of material from                                     & Count & 8.7k \\
\rowcolor{Gray}
nsfw       & Number of instances to tag all user posts as "Not Safe For Work"                                     & Count & 934 \\
\rowcolor{Gray}
media removal       & Number of instances to remove media from "                                     & Count & 630 \\
\rowcolor{Gray}
federated timeline removal       & Number of instances to un-list all user posts from the federated timeline  & Count & 2.4k  \\
\rowcolor{Gray}
posts\_tr       & Transformed number of posts using Box Cox transformation            & Count & 2.8k  \\
\rowcolor{Gray}
reject\_deletes      & Number of instances to remove all banners from       & Count & 158  \\
\rowcolor{Gray}
quaran\_inst      & Number of instances where private (DMs, followers-only) activities will not be sent     & Count & 1k  \\

\rowcolor{Gray}
mentions\_count       & Number of mentions in user posts on an instance                                      & Count & 24m \\
\rowcolor{Gray}
hate\_avg          & Average number of hate words on an instance from \url{hatebase.org}                             & Count  & 1.5   \\
\rowcolor{Gray}
url\_avg            & Average number of URLs in user posts on an instance                                            &  Count  & 0.2 \\
\rowcolor{Gray}
hashtags\_avg       & Average number of hashtags in user posts on an instance                                      & Count &  0.3 \\
\rowcolor{Gray}
mentions\_avg       & Average number of mentions in user posts on an instance                                      & Count & 0.8 \\
hashtags\_count       & Number of hashtags in user posts on an instance                                      & Count & 7m  \\

hate\_percent           & Average percentage of hate words in a post from \url{hatebase.org}                             & Percentage   &  2.2\%  \\
url\_percent            & Average percentage of URLs in user posts on an instance         &  Percentage &  8.4\% \\
hashtags\_percent       & Average percentage of hashtags in user posts on an instance      & Percentage & 6.6\%  \\
mentions\_percent       & Average percentage of mentions in user posts on an instance                                      & Percentage  & 2.5\%  \\

followers      & Number of followers of users on an instance                                      & Count & 169k \\
following       & Number of remote users that users on an instance follow                                     & Count & 8.9k \\
reblogs\_count       & Number of reblogs by users on an instance                                      & Count  & 7.2k \\
replies\_count       & Number of posts replied by users on an instance      & Count &  24.5k \\
users\_tr       & Transformed number of users using Box Cox transformation           & Count & 1.3k  \\

hate\_tr       & Transformed number of hate\_count using Box Cox transformation            & Count & 4.3k  \\
url\_tr       & Transformed number of url\_count using Box Cox transformation        & Count & 3k  \\

accept       & Number of instances to accept all material from       & Count & 635  \\
report removal      & Number of instances to remove all reports from         & Count & 91  \\
avatar removal       & Number of instances to remove all avatars from                  & Count & 266  \\
banner removal      & Number of instances to remove all banners from       & Count & 291  \\

followers\_only      & Number of instances that user posts are only seen by their followers         & Count & 99  \\

active\_halfyear      & Number of active users in half a year             & Count & 9  \\
active\_month      & Number of active users in a month                  & Count  & 7  \\

hash\_ftr      & Number of hashtags to remove activities from the federated timeline       & Count & 7  \\
hash\_rej      & Number of of hashtags to reject activities from         & Count & 6  \\
hash\_sen     & Number of  hashtags to mark activities as sensitive            & Count & 365  \\

\bottomrule
\end{tabular}}

\caption{Summary of all extracted features used for model training.}

 \label{table:features}
\end{table*}

\end{document}